\documentclass[%
 reprint,
superscriptaddress,
 preprintnumbers,
nofootinbib,
 amsmath,amssymb,
 aps,
]{revtex4-2}

\usepackage{graphicx}
\usepackage{dcolumn}
\usepackage{bm}
\usepackage{array}
\usepackage{hyperref}
\usepackage{orcidlink}
\usepackage{comment}
\usepackage{color}

\begin{document}

\preprint{ICPP--111}

\title{Radiative Signature of New Scalar Boson Decays in the $m_{\ell \ell \gamma}$ Spectrum at the LHC}

\author{Pramod Sharma\orcidlink{0000-0001-6381-7876}}
\email{pramod.sharma@cern.ch}
\affiliation{School of Physics and Institute for Collider Particle Physics, University of the Witwatersrand, Johannesburg, Wits 2050, South Africa}  
\affiliation{SGTB Khalsa College, University of Delhi, Delhi, India-110007.}

\author{Arnav Chauhan\orcidlink{0009-0004-7407-371X}}
\email{arnav.chauhan@cern.ch}
\affiliation{School of Physics and Institute for Collider Particle Physics, University of the Witwatersrand, Johannesburg, Wits 2050, South Africa}  
\affiliation{SGTB Khalsa College, University of Delhi, Delhi, India-110007.}

\author{Mukesh Kumar\orcidlink{0000-0003-3681-1588}}
\email{mukesh.kumar@cern.ch}
\affiliation{School of Physics and Institute for Collider Particle Physics, University of the Witwatersrand, Johannesburg, Wits 2050, South Africa}

\author{Andreas Crivellin\orcidlink{0000-0002-6449-5845}}
\email{andreas.crivellin@cern.ch}
\affiliation{Universitat Autònoma de Barcelona, 08193 Bellaterra, Barcelona}

\affiliation{ICREA, Instituci\'o Catalana de Recerca i Estudis Avan\c{c}ats,\\
Passeig de Llu\'{\i}s Companys 23, 
08010 Barcelona, Spain}

\author{Sukanta Dutta\orcidlink{0000-0002-9925-2085}}
\email{sukanta.dutta@sgtbkhalsa.du.ac.in}
\affiliation{SGTB Khalsa College, University of Delhi, Delhi, India-110007.}
\affiliation{Delhi School of Analytics, Institution of Eminence, University of Delhi, Delhi, India-110007.} 

\author{Rachid Mazini\orcidlink{0000-0003-0954-0970}}
\email{rachid.mazini@cern.ch}
\affiliation{School of Physics and Institute for Collider Particle Physics, University of the Witwatersrand, Johannesburg, Wits 2050, South Africa}  

\author{Bruce Mellado\orcidlink{0000-0003-4838-1546}}
\email{bmellado@mail.cern.ch}                   
\affiliation{Institute of High Energy Physics, 19B, Yuquan Road, Shijing District, 100049, Beijing, China}
\affiliation{School of Physics and Institute for Collider Particle Physics, University of the Witwatersrand, Johannesburg, Wits 2050, South Africa}  

%

\date{\today}

\begin{abstract}

We investigate the radiative decay $S \to W^+W^-\gamma$ in the context of the multi-lepton anomalies and recent indications of a narrow scalar resonance near $m_S = 152 \pm 1~\text{GeV}$ in the $\gamma\gamma$, $Z\gamma$, and $W^+W^-$ channels at the Large Hadron Collider. These excesses arise in final states containing leptons, missing transverse momentum, and associated $b$-jets, and motivate a search for a corresponding localized excess in the invariant-mass spectrum of the dilepton--photon system, $m_{\ell\ell\gamma}$, in events with associated $b$-jets. We use recent CMS measurements of the ${t\bar{t}}\gamma$ differential cross sections~\cite{CMS:2025zbe} to study the $m_{\ell\ell\gamma}$ spectrum and perform a search for a scalar-resonance contribution. A localized excess is observed, compatible with the scalar-resonance hypothesis, with a global significance of $2.7\sigma$ at $m_S = 152~\text{GeV}$. This result provides additional support for the hypothesis of a narrow resonance. The ratio $\sigma(S \to W^+W^-\gamma)/\sigma(S \to W^+W^-) = (2.14 \pm 0.77)$\% is extracted. This value is compatible with an enhanced radiative contribution that could arise in scenarios beyond the Standard Model.

\end{abstract}

\maketitle


\section{Introduction}

The Standard Model (SM) of particle physics provides a highly successful theoretical framework for describing the fundamental constituents of matter and their interactions at microscopic scales~\cite{Weinberg:1967tq}.
Its predictions have been tested extensively over a wide range of energies~\cite{ParticleDataGroup:2026aaa}. A major milestone was reached in 2012 with the discovery of the Brout--Englert--Higgs boson, $h$, at the Large Hadron Collider (LHC) at CERN~\cite{ATLAS:2012yve,CMS:2012qbp}. In addition, the measurements show consistency with the quantum numbers and branching ratios of the 125~GeV Higgs boson with Standard Model expectations~\cite{CMS:2022dwd,ATLAS:2022vkf}.

Despite its remarkable success, the SM is clearly not the final theory of fundamental interactions. In particular, it does not account for dark matter inferred from astrophysical and cosmological observations~\cite{Planck:2018vyg}, nor for the non-zero neutrino masses implied by neutrino oscillation data~\cite{T2K:2025wet}. It also fails to explain the observed matter--anti-matter asymmetry of the Universe~\cite{Sakharov:1967dj,Planck:2018vyg}. In addition, several conceptual issues remain unresolved, including the flavour puzzle, the hierarchy problem, and the stability of the scalar potential. Since many of these open questions are closely connected to the scalar sector, extensions of the Higgs boson sector are especially well motivated. Such extensions can also provide viable links to dark matter, for example, through Higgs-portal scenarios~\cite{Arcadi:2019lka}. This makes the Higgs sector a particularly promising place in which to search for physics beyond the Standard Model (BSM).

A wide range of searches for new particles in general, and in particular for additional scalar states, has been carried out at the LHC. Although these dedicated searches have not produced conclusive evidence for new resonances in the energy range explored so far, the so-called multi-lepton anomalies have emerged. These consist of tensions in final states containing multiple leptons (electrons and/or muons), together with missing transverse energy and with or without $b\text{-}$jets~\cite{vonBuddenbrock:2017gvy,Buddenbrock:2019tua,Hernandez:2019geu,vonBuddenbrock:2020ter}; see Refs.~\cite{Fischer:2021sqw,Crivellin:2023zui} for reviews and further references. The corresponding discrepancies are statistically very significant, exceeding the $5\sigma$ level, and point to the associated production of new electroweak-scale scalars~\cite{vonBuddenbrock:2016rmr,Buddenbrock:2019tua}. In particular, within a simplified-model framework, the multi-lepton anomalies are compatible with the assumptions of the production\footnote{The dominant process is $pp \to H \to SS^\prime$ via gluon-fusion. From this point onward, we use a shorthand notation in which only the decay of $S$ is explicitly stated, with the production mechanism implicitly understood to be gluon fusion followed by the decay of the heavy scalar $H$. For example, $\sigma(S \to W^+W^-)$ denotes the cross section for the process $gg \to H \to SS^\prime$, with $S \to W^+W^-$ and, for instance, $S^\prime \to b\bar{b}$.} via gluon fusion of a scalar $H$ with a mass of approximately $m_H=270$~GeV, which decays predominantly into a pair of lighter scalars $S$, one of them off shell, that in turn decay mainly into $W$ bosons. For $S$, a mass of $150\pm5$\,GeV was inferred from the invariant-mass distribution of an electron--muon pair arising from resonant $WW$ signals~\cite{vonBuddenbrock:2017gvy}.

\begin{figure}[t]
    \centering
    \includegraphics[width=0.8\linewidth]{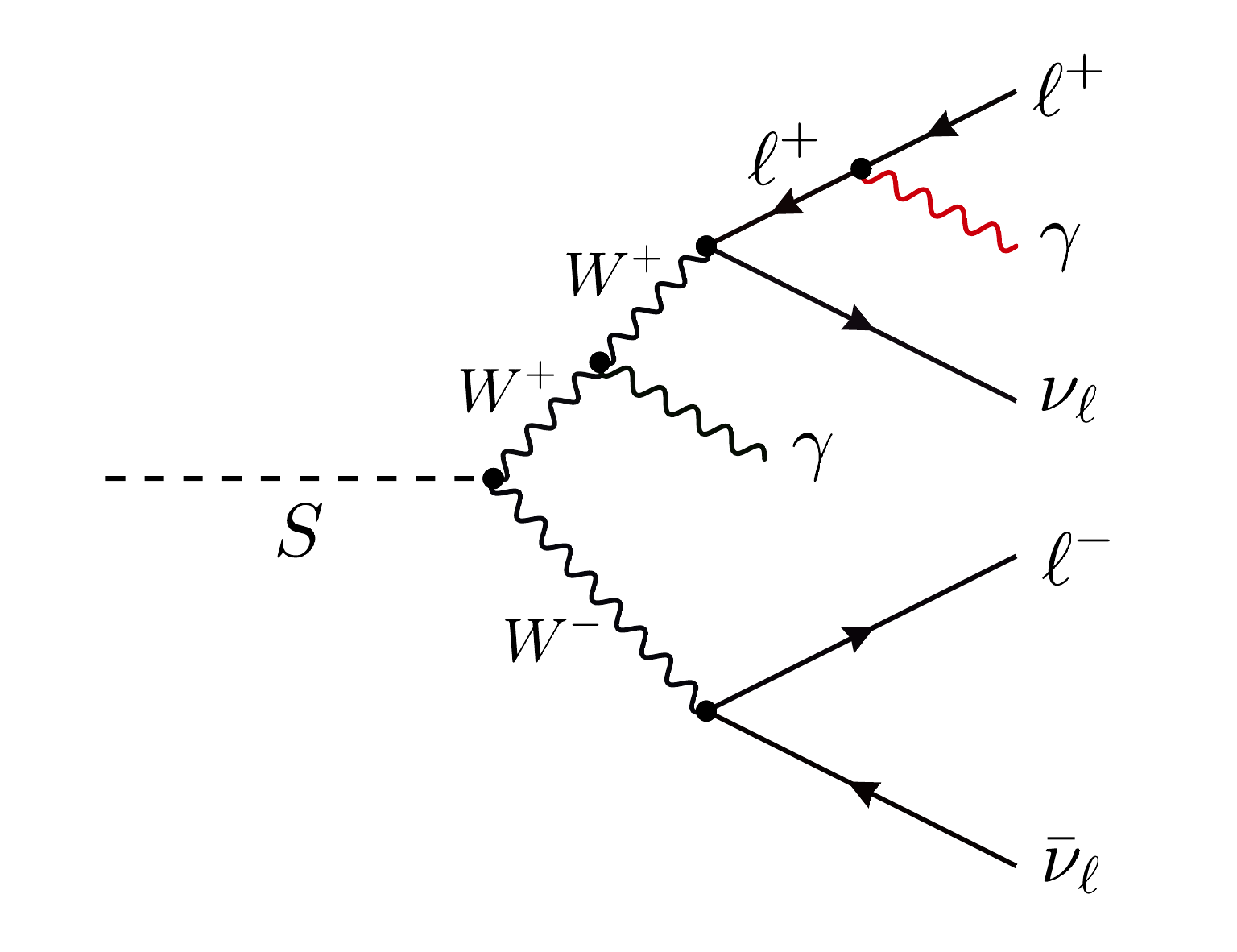}
    \caption{Feynman diagram showing the decay  $S \to \ell^+ \nu_\ell \, \ell^- \bar{\nu}_\ell \, \gamma \, (\ell = e, \mu, \tau)$. The photon is radiated from the positively charged $W^+$ boson or the $\ell^+$; analogous contributions arise from radiation off the $W^-$ boson and $\ell^-$.}
    \label{fig:feynmandiag}
\end{figure}

This motivates searches for $S$ in resonant channels in the same mass region, in particular in $S\rightarrow\gamma\gamma$, $Z\gamma$, $W^+W^-$, and $ZZ$ channels. Indeed, Refs.~\cite{Crivellin:2021ubm,Coloretti:2023wng,Bhattacharya:2025rfr} showed that the sidebands of LHC analyses $h \rightarrow \gamma\gamma,Z\gamma$, $W^+W^-$, where $h$ is the SM Higgs boson, performed by ATLAS~\cite{ATLAS:2021jbf,ATLAS:2023omk,ATLAS:2022ooq} and CMS~\cite{CMS:2018myz,CMS:2022ahq,CMS:2022uhn} yield indications for a narrow resonance excess with a mass $152\pm1$\,GeV. This particle is produced in association with leptons, missing energy, $b$-jets, and electroweak gauge bosons, as inferred from the interpretation of the multi-lepton anomalies. The global significance of this narrow resonance excess surpasses 5$\sigma$ after the application of residual look-elsewhere effects. The absence of a signal in the $S\rightarrow ZZ$ suggests that $S$ may be the neutral component of a triplet with hypercharge 0, as proposed in Refs.~\cite{FileviezPerez:2008bj,Coloretti:2023yyq}.

In this letter, we examine a new type of multi-lepton excess that has emerged with the Run-2 data at the LHC, which involves the production of two opposite-sign leptons $(\ell=e,\mu)$, a photon, and at least a $b$-jet. This excess could be interpreted as a radiative correction to $S \rightarrow WW$, i.e.,  $S\rightarrow W^+W^-\gamma$ (see Figure~\ref{fig:feynmandiag}) produced in association with $b$-jets, as suggested by the multi-lepton anomalies. The significance of the excess is obtained using the simplified model $H\rightarrow SS^\prime$, where $H$ is a heavy scalar with $m_H=270$\,GeV and $S^\prime$ is a lighter scalar of mass $m_{S^\prime}=95$\,GeV, which can be used to explain the multi-lepton anomalies~\cite{vonBuddenbrock:2016rmr,Buddenbrock:2019tua,Coloretti:2023yyq}. We vary the mass of the scalar $S$, $m_S$, from 100\,GeV to 200\,GeV. In addition to the significance of this new excess, we extract the preferred value of the mass and the ratio of the observed cross-sections $\sigma(S\rightarrow W^+W^-\gamma) /\sigma(S\rightarrow W^+W^-)$ in association with at least one $b$-jet.

\section{Data set and methods}
\label{dataset}

The CMS Collaboration~\cite{CMS:2025zbe} has recently presented inclusive and differential measurements of the ${t\bar{t}}\gamma$ production cross section and the cross-section ratio $\sigma({t\bar{t}}\gamma)/\sigma({t\bar{t}})$. In particular, it reported the distribution of the invariant mass of the dilepton--photon system, $m_{\ell\ell\gamma}$ (see Figure~4, right, in Ref.~\cite{CMS:2025zbe}), which had not been provided in previous studies of this final state~\cite{ATLAS:2015jos,ATLAS:2017yax,CMS:2017tzb,CMS:2021klw,ATLAS:2018sos,ATLAS:2024hmk,ATLAS:2022wec}. The $m_{\ell\ell\gamma}$ distribution reported by CMS exhibits good agreement between the observed and predicted event yields for $m_{\ell\ell\gamma}>120$\,GeV, as indicated by the data-to-prediction ratio shown in the lower panel. We therefore use this region to determine an overall normalization of the SM background prediction. Specifically, the background is normalized to the data using a factor defined as the ratio of the total number of observed events over the total predicted SM background yield above 120\,GeV. The resulting normalization factor is 0.98, consistent with unity within the quoted systematic uncertainties.

\begin{figure}[t]
    \centering
    \includegraphics[width=1\linewidth]{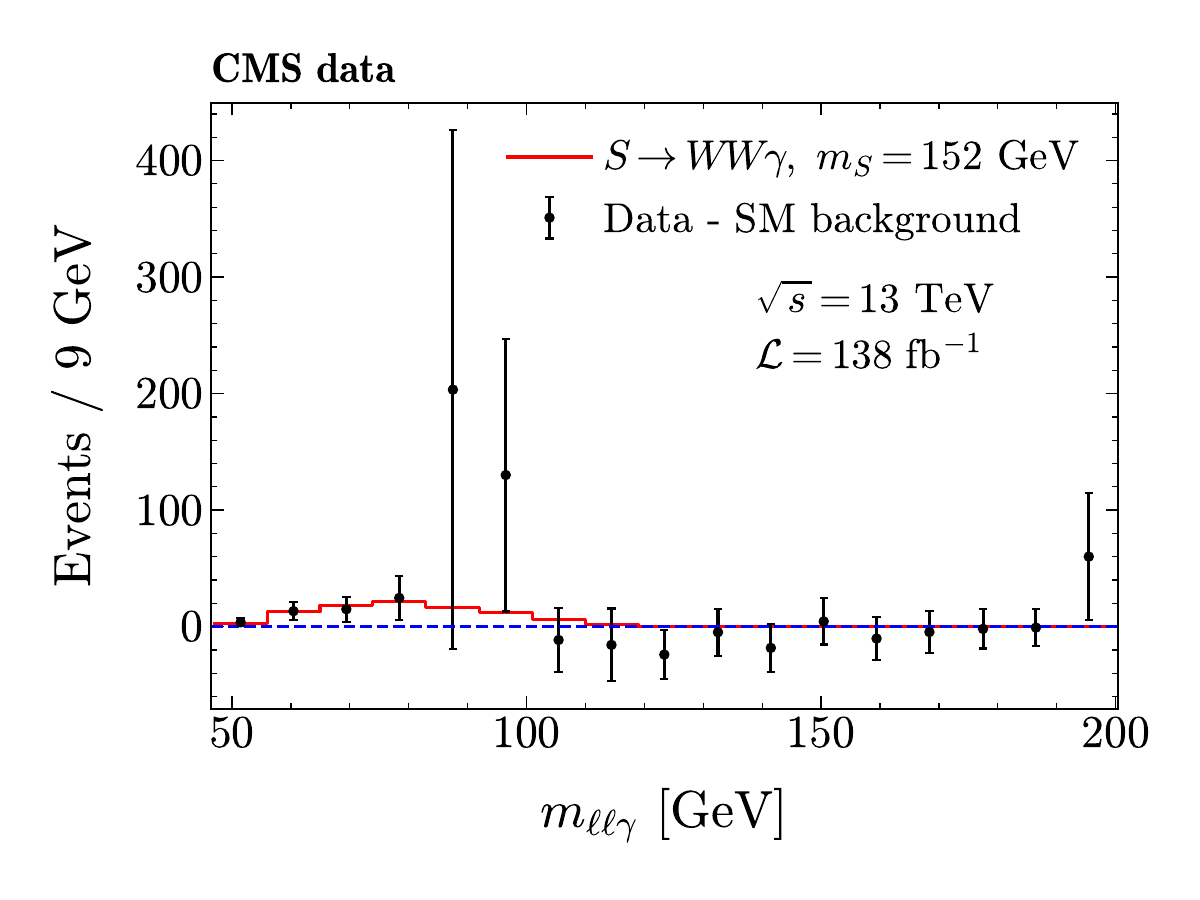}
    \caption{Event yields per 9~GeV after subtraction of the normalized SM backgrounds as a function of $m_{\ell\ell\gamma}$ in GeV. Here, {\tt{Data $-$ SM background}} is defined as ``excess." The background is normalized by a scale factor derived from the ratio of total event yields in data and background in the region $m_{\ell\ell\gamma} > 120 $\,GeV (see text). The BSM signal for $m_S = 152$~GeV is represented by a red line.}
    \label{fig:excess}
\end{figure}

Figure~\ref{fig:excess} displays the event yields after subtraction of the normalized SM backgrounds as a function of $m_{\ell\ell\gamma}$. The dominant backgrounds to the BSM signal are the production of $t\overline{t}$ in association with a photon and non-prompt backgrounds. The error bars contain the statistical and systematic errors combined in quadrature. The large errors around the $Z$-boson mass are due to the large systematic uncertainties of the theory prediction of the $Z\gamma$ backgrounds. It is evident that the data display an excess within a region just below 110\,GeV, as one would expect from a resonant decay of $S\rightarrow W^+W^-\gamma$.

For the purposes of the present study, both the SM process: $pp \rightarrow t \bar{t} \rightarrow \ell^+ \nu_\ell \ell^- \bar{\nu}_\ell \gamma b \bar{b}$ and the BSM process: $pp \rightarrow H \rightarrow SS^\prime$, with $S \rightarrow \ell^+ \nu_\ell \ell^- \bar{\nu}_\ell \gamma$ (as shown in Figure~\ref{fig:feynmandiag}) and $S^\prime \rightarrow b \bar{b}$, are simulated using {\tt MadGraph5}~\cite{Alwall:2011uj} for the matrix-element event generation. The resulting parton-level events are then processed through the standard simulation chain, consisting of {\tt Pythia 8.3}~\cite{Sjostrand:2014zea} for particle decays, parton showering, initial- and final-state radiation, fragmentation, and hadronization, followed by {\tt Delphes 3.5.0}~\cite{deFavereau:2013fsa} for detector simulation and particle-level object reconstruction.

The event selection, taken from Ref.~\cite{CMS:2025zbe}, is defined as follows: Each event must contain at least two charged leptons ($\ell=e,\mu$). Electron candidates are selected if $p_T > 25$\,GeV for the leading lepton and $p_T > 20$\,GeV for the sub-leading leptons within the $|\eta| < 2.5$ region. Muons are required to satisfy $p_T > 25$\,GeV for the leading lepton and $p_T > 15$\,GeV for the sub-leading lepton, and $|\eta| < 2.4$. The two highest $p_T$ leptons are selected as the leading pair and are required to have opposite electric charges. The allowed dilepton final states are $e^\pm\mu^\mp$, $e^\pm e^\mp$, and $\mu^\pm\mu^\mp$. The pair of leptons must have an invariant mass $m_{\ell \ell} > 40$\,GeV.  In addition, exactly one photon is required in the event with $p_T > 20$\,GeV and must lie within $|\eta| < 2.5$. Jets are required to have $p_T > 30$\,GeV and $|\eta| < 2.4$. A minimum of two jets is required, with at least one of them identified as a $b$-tagged jet.

A correction is applied to account for residual detector effects not fully captured by the fast simulation. This correction is derived using the $t\bar{t}\gamma$ process, 
$pp \rightarrow t \bar{t} \rightarrow \ell^+ \nu_\ell \ell^- \bar{\nu}_\ell \gamma b \bar{b}$, as a calibration channel. 
For this process, the selection efficiency obtained from the Monte Carlo (MC) simulation is compared with that inferred from Ref.~\cite{CMS:2025zbe}. The ratio of these efficiencies is 1.01, which is used as a correction factor to the BSM signal process.

\begin{figure}[t]
    \centering
\includegraphics[width=1\linewidth]{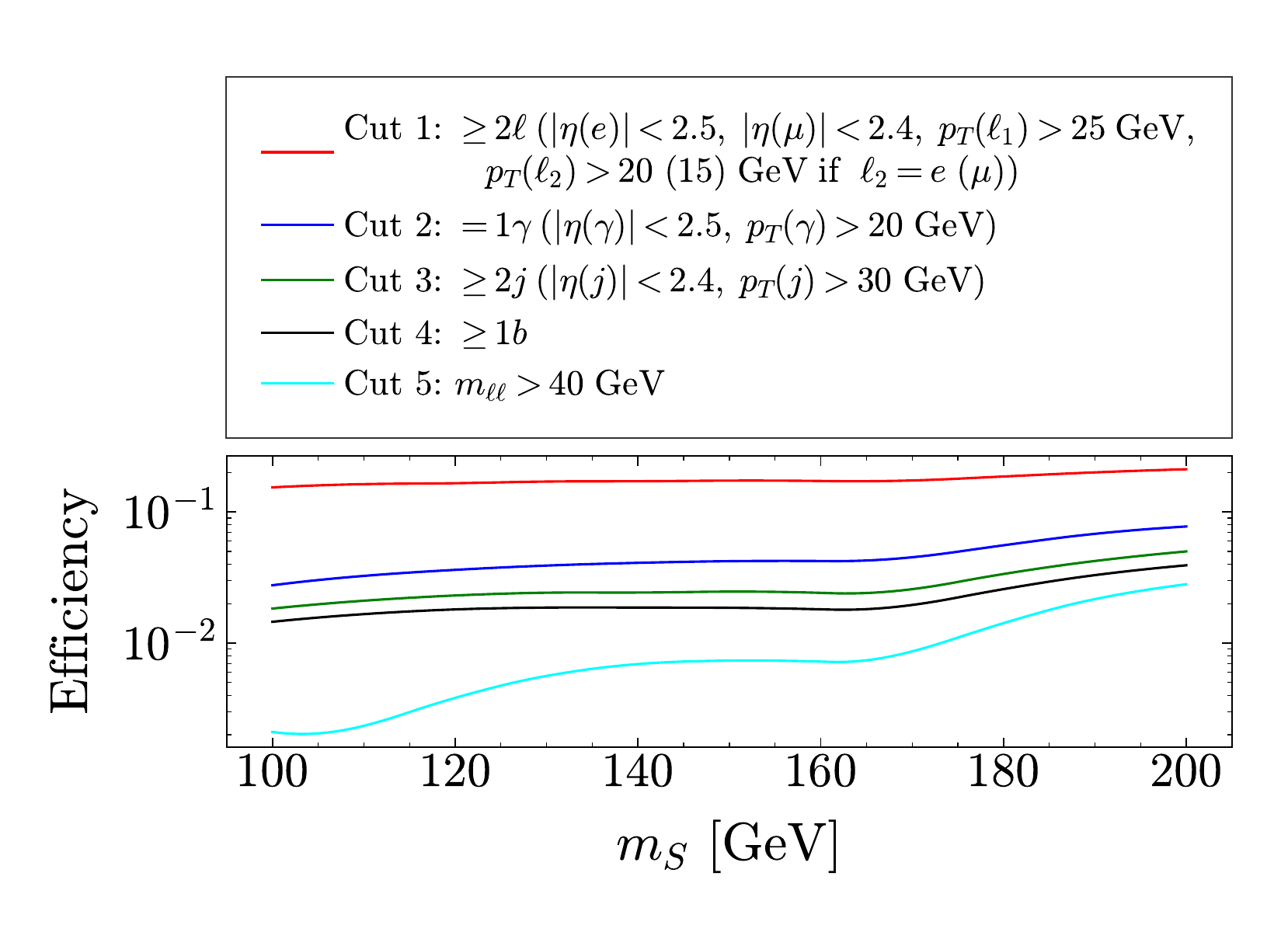}
    \caption{Selection efficiencies (see text) for mass values $m_S \in [100,200]$. Here, $\ell_1$ and $\ell_2$ are leading and sub-leading leptons, respectively. }
    \label{fig:placeholder}
\end{figure}

Figure~\ref{fig:placeholder} reports the efficiency of the BSM signal after applying the correction factor of 1.01. 
The efficiencies are organized in terms of four consecutive steps: leptonic, photonic, jet, and $b$-jet requirements. Finally, the efficiencies of the last requirement, $m_{\ell\ell}>40$\,GeV, are also reported. 

Recent reports of excess events in the vicinity of the $t\overline{t}$ threshold by CMS and ATLAS~\cite{CMS:2025kzt,ATLAS:2026dbe} have reignited theoretical discussion about the adequacy of QCD predictions in this region. In particular, the claim that the data can be described by adding a color-singlet pseudoscalar $\eta_t$ resonance, or ``toponium,'' on top of the standard fixed-order perturbative-QCD MC predictions remains controversial. At the same time, these studies underscore that further work is still required to incorporate threshold effects consistently into MC simulations~\cite{Nason:2025hix,Flacke:2025dwk}. To determine the possible impact of a toponium contribution on the $m_{\ell\ell\gamma}$ spectrum, we generate the process ($pp\to \eta_t \to t\bar t\gamma$). Here, we have adopted a conservative approach by assuming the cross-section of $8.8^{+1.2}_{-1.4}$\,pb~\cite{CMS:2025kzt}. After applying the same event selection cuts defined above in this section, the corresponding event yields from the $\eta_t$ contribution are 25 and 15 in the regions $m_{\ell\ell\gamma} < 120$~GeV and $m_{\ell\ell\gamma} > 120$~GeV, respectively. These estimates, while likely an overestimate of the $t\overline{t}$ threshold effects, yield a very small contribution, and this is unable to explain the observed excess. In addition, the model used here provides a broad $m_{\ell\ell\gamma}$ distribution that is incompatible with the shape of the observed excess. 

\section{Results}
\label{results}

We extract the mass of the scalar particle $S$, with $m_{S} \in [100,\,200]~\text{GeV}$, by minimizing the $\chi^2$ constructed from 8 bins of the $m_{\ell\ell\gamma}$ distribution measured in~\cite{CMS:2025zbe} for $m_{\ell\ell\gamma} < 120~\text{GeV}$, where $\chi^2(m_S)$ is defined as: 
\begin{align}
\chi^2(m_S)
&=
\sum_{i=1}^{8}
\frac{
\left[
E_i - N^{\mathrm{BSM}}_i(m_{S})
\right]^2
}{
\sigma_i^2
},\notag
\\[6pt]
E_i
&=
N^{\mathrm{data}}_i - N^{\mathrm{SM\,bkg}}_i \,. 
\end{align}
Here, $i$ denotes the bin number. $N^{\mathrm{data}}_i$ is the number of observed events in the $i$-th bin, while $N^{\mathrm{SM\,bkg}}_i$ represents the SM background prediction corrected by the normalization factor discussed in Sec.~\ref{dataset}. The quantity $E_i$ is defined as the bin-wise difference between the observed data and the SM background prediction. $N^{\mathrm{BSM}}_i(m_S)$ denotes the BSM signal contribution. {The total uncertainty in the $i$-th bin, represented by $\sigma_i$, is defined as
\begin{equation}
\sigma_i^2 =
\left(\sigma_i^{\mathrm{stat}}\right)^2 +
\left(\sigma_i^{\mathrm{syst}}\right)^2 ,
\end{equation}
where $\sigma_i^{\mathrm{stat}}$ and $\sigma_i^{\mathrm{syst}}$ denote the statistical and systematic uncertainties, respectively. The most relevant systematic uncertainty is associated with the $Z\gamma$ background. It is estimated from the average relative uncertainty originating from the $Z\gamma$ background in the two dominant bins of the $m_{\ell\ell\gamma}$ distribution, namely the 5$^{\rm th}$ and 6$^{\rm th}$ bins. This average relative uncertainty is then applied to all bins to construct $\sigma_i^{\mathrm{syst}}$. Note that we adopt a conservative approach, where systematic uncertainties are assumed to be uncorrelated between bins.

\begin{figure}[t]
    \centering
\includegraphics[width=1.0\linewidth]{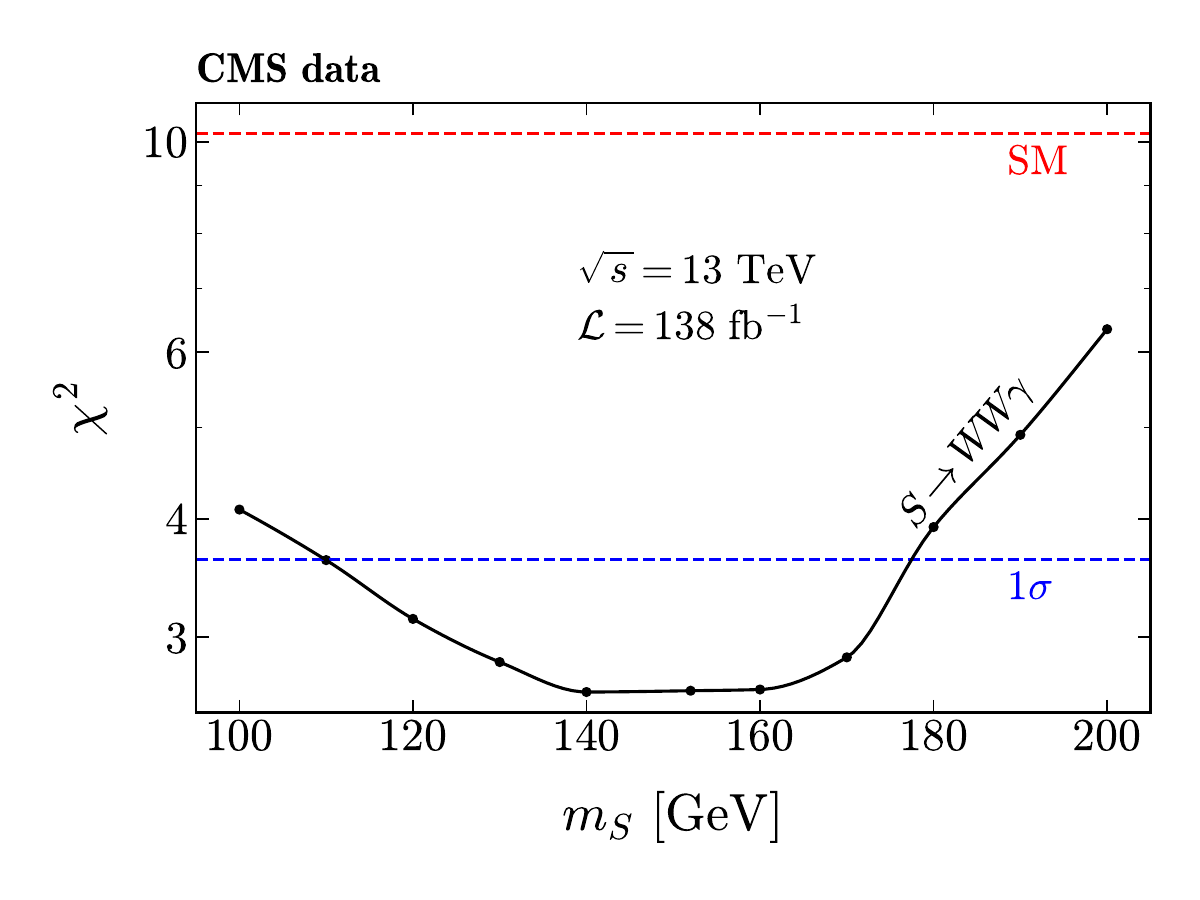}
    \caption{$\chi^2$ as a function of the scalar mass $m_S$ (see text for details).}
    \label{fig:ChiVsmS}
\end{figure}

Figure~\ref{fig:ChiVsmS} presents the $\chi^2$ distribution as a function of the scalar mass $m_S$. The minimum occurs at $m_S=140$\,GeV, where $\chi^2=2.63$, while the no-BSM hypothesis yields $\chi^2=10.21$. At the 68\% confidence level, the constraint on the scalar mass is $m_S=140^{+35}_{-30}$\,GeV. The relatively wide confidence interval is primarily a consequence of the restricted phase space imposed by the requirement $m_{\ell\ell}>40\,\mathrm{GeV}$, compounded by the large systematic uncertainties near the $Z$-boson mass. The small value of the minimum $\chi^2$ could be due to an overestimation of uncertainties. Evaluated at $m_S=152\,\mathrm{GeV}$, the fit yields $\chi^2=2.63$, corresponding to a significance of $2.7\sigma$.

We use the results of Ref.~\cite{Buddenbrock:2019tua}, which extracted the $S\to W^+W^-$ signal cross-section for a scalar mass of $m_S=150\,\mathrm{GeV}$ using opposite-sign dilepton data reported by the CMS Collaboration with Run-2 data~\cite{CMS:2018amb}. At the same mass value, we estimate the cross-section for  $S \to W^+W^-\gamma$ to be $0.92\pm0.33,\mathrm{pb}$. Combining the two results, we obtain
\begin{align*}
    \frac{\sigma(S \to W^{+}W^{-}\gamma)}{\sigma(S \to W^{+}W^{-})} = (2.14 \pm 0.77)\% .
\end{align*}
This value is somewhat larger than the expectation from purely SM radiative processes, which is of order 1\%, and may therefore point to an enhancement driven by physics beyond the Standard Model. However, more data will be needed to shed light on this hint. 

\section{Conclusions}
\label{conc}

\begin{table}[t] 
\centering
\renewcommand{\arraystretch}{2.3}
\begin{tabular}{c|c|c} 
\hline
Mass (GeV) & Channel & Reference \\ 
\hline 
$150^{+5}_{-5}$ & \shortstack{$S\rightarrow W^+W^-\rightarrow \ell^+\nu_\ell\ell^-\nu_\ell$ \\($b$-jet veto applied)} &\cite{vonBuddenbrock:2017gvy} \\ 
$150^{+7}_{-6}$ & \shortstack{$S\rightarrow W^+W^-\rightarrow \ell^+\nu_\ell\ell^-\nu_\ell$ \\ (with associated $b$-jet)}& \cite{Banik:2023vxa}\\
$152^{+1}_{-1}$ & $S\rightarrow \gamma\gamma, Z\gamma$ & \cite{Crivellin:2021ubm,Bhattacharya:2025rfr} \\
$140^{+35}_{-30}$ & $S\rightarrow W^+ W^- \gamma\rightarrow \ell^+\nu_\ell\ell^-\nu_\ell \gamma$ & This work \\
\hline
\end{tabular} 
\caption{Mass measurements of a scalar candidate around 150\,GeV (see text).}
\label{tab:comparison} 
\end{table}

In this Letter, we investigate the excess observed in the $\ell^+\ell^-\gamma$ final state events, with $\ell=e,\mu$, produced in association with a $b$-jet in the low-$m_{\ell\ell \gamma}$ region. We interpret this using a simplified BSM framework that is compatible with the multi-lepton anomalies reported at the LHC and with earlier indications of a narrow resonance near $152$ GeV~\cite{vonBuddenbrock:2017gvy,Buddenbrock:2019tua,Hernandez:2019geu,vonBuddenbrock:2020ter,Fischer:2021sqw,Crivellin:2023zui}. We find a significance of $2.7\sigma$ at $m_S = 152$ GeV, together with a preferred mass range of $140^{+35}_{-30}$ GeV, consistent with earlier estimates~\cite{vonBuddenbrock:2017gvy,Crivellin:2021ubm,Bhattacharya:2025rfr}. Table~\ref{tab:comparison} displays the comparison of the available mass measurements of the scalar candidate.\footnote{The preferred scalar mass reported in Ref.~\cite{Banik:2023vxa} could be affected by possible unaccounted-for $t\overline{t}$ threshold effects. Such effects may have an impact on the mass measurement and should be kept in mind when making a direct comparison with our result.} This result provides additional support for the hypothesis of a narrow resonance near 152\,GeV.

We also determine the ratio $\sigma(S \to W^{+}W^{-}\gamma)/\sigma(S \to W^{+}W^{-}) = 2.14 \pm 0.77\%$, which is larger than expected from radiative corrections induced by SM processes alone. Radiative corrections to direct $S\to W^+W^-$ production are known in related contexts~\cite{Kniehl:1991xe,Bredenstein:2006rh,Dawson:2018liq}, whereas possible enhancements from BSM virtual effects have not been studied extensively. Nevertheless, there are strong indications that extended Higgs sectors can give rise to particularly large electroweak corrections~\cite{Egle:2023pbm}. Another possible interpretation is an accidental overlap with the decay of a charged scalar into $W^\pm\gamma$ in association with an additional lepton~\cite{Degrande:2017naf}.


The dedicated analyses of the $t\bar t\gamma/t\bar t$ ratio presented in Refs.~\cite{Bevilacqua:2018woc,Bevilacqua:2019quz} show that this observable benefits from strong correlations between the two processes, leading to a substantial reduction of common theoretical uncertainties. In the present study, we adopt a conservative treatment of possible $t\bar t$ threshold effects and find that they are too small to account for the observed differential excess.




On the experimental side, this Letter supports relaxing the $m_{\ell\ell}>40$\,GeV requirement in future measurements of the $t\bar{t}\gamma$ cross section and of the differential $t\bar{t}\gamma/t\bar{t}$ cross-section ratio. Extending the accessible $m_{\ell\ell\gamma}$ range, together with the larger Run-3 data set, will be important for clarifying the nature and detailed shape of the excess.

\begin{acknowledgments}
The authors would like to thank Michael Ramsey-Musolf for invaluable discussions. BM would like to thank the Institute of high Energy Physics of Chinese Academy of Sciences for their invaluable support. PS and SD acknowledge partial support from the Anusandhan National Research Foundation (ANRF), Government of India, through Grant No. CRG/2023/008234. The authors would like to acknowledge the support from the Department of Science, Technology, and Innovation through the SA-CERN program and other forms of support. 

\end{acknowledgments}



\nocite{*}

\bibliography{ref}

\end{document}